\documentclass[runningheads]{llncs}

\usepackage{graphicx}
\usepackage{amsmath,amssymb}
\usepackage{multirow}

\begin{document}

\title{Public Opinion as an Option: Leveraging Prediction Markets to Hedge Exposure to Spot Crypto Volatility}
\titlerunning{Public Opinion as an Option}

\author{Prashanth Bhaskara\quad Aadit Jerfy}
\authorrunning{Bhaskara, Jerfy}

\institute{The University of Chicago\\
pbhaskara@uchicago.edu, ajerfy@uchicago.edu}

\maketitle

\begin{abstract}
This paper proposes an investment strategy through resource allocation into Kalshi Crypto Event Contracts in order to effectively hedge exposure to spot asset volatility. Using Bitcoin as a proof of concept, we treat corresponding Kalshi markets on the asset’s future price as option contracts, and through construction of different portfolio allocations present a framework for which event contracts can be effectively utilized by retail investors in order to control risk framework without reducing underlying exposure to spot volatility. We showcase this information through analysis of three separate Kalshi event contracts, all relating to a different underlying market regime (i.e bullish, bearish, flat), and provide a comprehensive risk profile for each regime. Results show that a dynamic Kalshi/Bitcoin hedging strategy outperforms more simplistic portfolios across all regimes, supporting our initial hypothesis. We discuss further considerations which can erode profits in the future (i.e. fees), and provide direction for future research.
\keywords{prediction markets \and hedging \and cryptocurrency \and Kalshi \and risk management}
\end{abstract}

\section{Introduction}
Cryptocurrencies have consistently remained a difficult commodity to effectively
hedge against for retail traders due to token volatility. Traditional hedging measures, including
futures and contract for differences, are hampered by basis risks, margin requirements, and lack
of access for retail traders in the United States [1]. Option contracts, while ideal, suffer from high
premiums and low liquidity, and aren’t realistic means for hedging exposure either. Beginning in
early 2024, Kalshi launched prediction market events centered around allowing users to trade on
future prices of various cryptocurrencies. Such contracts are readily available and aggregate
public opinion in order to generate a derived forecast of the spot asset, therefore serving as
potential risk diversifying instruments. In this paper, we propose treating such markets as
leveraged hedging instruments on the spot asset, denoting the event as an interchangeable call
option in order to effectively mitigate exposure during volatile circumstances. Using a historical
dataset of Kalshi event contracts on Bitcoin’s price during distinct market regimes, we
construct a comprehensive risk profile of a two-fund portfolio consisting of the spot asset
and Kalshi contracts, rebalanced periodically in order to reflect evolving market
conditions. We then compare the constructed portfolio to both a simplistic buy-and-hold
strategy and well as a static hedge in order to demonstrate Kalshi’s effectiveness as an
effective financial instrument for all traders in the digital assets industry.

\section{Methodology}

\subsection{Data}
$\boldsymbol{BTC}$ $\boldsymbol{Spot}$ $\boldsymbol{Price}$: 1-minute OHLCV data for the BTC/USDT trading pair from the Binance API, encompassing 2024, 2025, and January of 2026. 
\\
\\
$\boldsymbol{Kalshi}$ $\boldsymbol{Market}$ $\boldsymbol{Data}$ :1-minute-level pricing data for binary outcome contracts on BTC price targets from Kalshi:
\begin{center}
    \begin{itemize}
        \item "Will BTC max reach \$100k+ by Dec 31 2024? [2]"
        \item "Will BTC close above \$100k by December 31 2025? [3]"
        \item "Will BTC max reach \$100k+ by Jan 31 2026? [4]"
    \end{itemize}
\end{center}
 These markets were open from March - December (2024), November - December (2025), and January (2026) respectively. The three tested periods represent distinct market regimes, with 2024 capturing a bullish movement (36\% BTC return), December 2025 capturing moderate volatility and sideways price action (-3.6\%), and January 2026 capturing a bearish drawdown (-26\%).

\subsection{Probability modeling framework}
We employed a two-component estimation approach to identify differences in market sentiment and theoretical probabilities. The market-implied probability is determined by Kalshi contract prices (i.e., a contract trading at 65 cents implies that the market assigns a 65\% probability to the “Yes” outcome). These probabilities were extracted and converted to decimal probabilities. 
\\\\Treating the market question’s fundamental price level as a strike price, we utilize a modified formulation of the Black-Scholes model in order to calculate the theoretical probability the asset expires “In the Money”. We restrict our derivation to risk-neutral-probabilities exclusively, in consideration of Bitcoin’s lack of a risk-free-benchmark. Annualized volatilities are calculated monthly for BTC, with each event contract utilizing BTC’s volatility the month prior to the market’s open. Additionally, considering our event contracts reflect the probability the maximum value of the asset reaches the strike price, we immediately default our probabilities to 1 if the asset’s price ever breaches the target price level. 

\newpage \noindent More formally,

\begin{equation}
p_i =
\begin{cases}
1, & \text{if } \exists j \le i \text{ such that } \mathrm{BTC}_j \ge K,\\[4pt]
\Phi\!\left(\dfrac{\ln(S_i/K) - \tfrac{1}{2}\sigma^2 T}{\sigma \sqrt{T}}\right), & \text{otherwise.}
\end{cases}
\label{eq:pi_piecewise}
\end{equation}
\\
where $p_i$ equals the probability BTC expires ITM at time i, $K$ equals the strike price provided by the Kalshi event contract, $S$ = BTC’s price at time i, $\sigma^2$ = annualized historical volatility for BTC,  $T$ = time to expiry for event contract, $\Phi$ = the normal distribution CDF, and $BTC_j$ equals BTC’s price at time j.
\\\\
Note that we choose to utilize a modified vanilla style Black-Scholes model over a barrier one-touch derivation since our objective is to identify relative mispricing, and the vanilla model does so without the increased risk brought on by raised parameter sensitivity. Moreover, limited liquidity impedes continuous price adjustment following threshold breaches (i.e., the contracts being traded wouldn't represent exact barrier payoffs as barrier pricing models assume).

\subsection{Trading strategy}
We define mispricing as the difference between the theoretical and market-implied probabilities, $D = p(\mathrm{Actual}) - p(\mathrm{Implied})$, and implement a five-level signal system:
\begin{center}
\begin{itemize}
  \item \textbf{+2}: $D > 15\%$ (strong buy), 5000 contracts
  \item \textbf{+1}: $D > 5\%$ (weak buy), 3000 contracts
  \item \textbf{+0}: $|D| < 2\%$ (neutral), close position
  \item \textbf{ -1}: $D < -5\%$ (weak short), $-3000$ contracts
  \item \textbf{ -2}: $D < -15\%$ (strong short), $-5000$ contracts
\end{itemize}
\end{center}

\noindent Note that a short equates to buying the \textit{NO} contract for the event.
\\\\
Three distinct strategies were constructed (each with \$100k initial capital).
\begin{center}
\begin{itemize}
  \setlength{\itemsep}{0.8em}  
  \setlength{\parsep}{0.2em}   

  \item \textbf{Strategy A} - Unhedged buy / hold BTC spot position.
  \item \textbf{Strategy B} - Static hedge, combining a long BTC spot and single short position (1000 contracts) on Kalshi “Yes” contracts entered at the beginning of the period and held to expiry.

  \item \textbf{Strategy C} - Combined position of long BTC spot and active trading of Kalshi contracts based on the pricing signals outlined above.
\end{itemize}
\end{center}
\vspace{1cm}
\noindent For Strategy C, the  portfolio is rebalanced at variable intervals (each month, week, day, 12 hrs, 6 hrs, 1hr) to limit the number of transactions (and subsequently minimize fee impact). The signal is sampled at each of these fixed horizons, and contract positions are taken when pricing differentials are detected. Note that the strategies do not take into account latency and market liquidity concerns, creating potential deviations from our outputs and real-world outcomes. However, we posit that the  opportunities Kalshi provides within these spaces will provide incentives for these markets to receive more institutional/retail adoption, ultimately mitigating concerns and converging results.

\begin{figure}[htbp]
  \centering
  \includegraphics[width=1.0\textwidth]{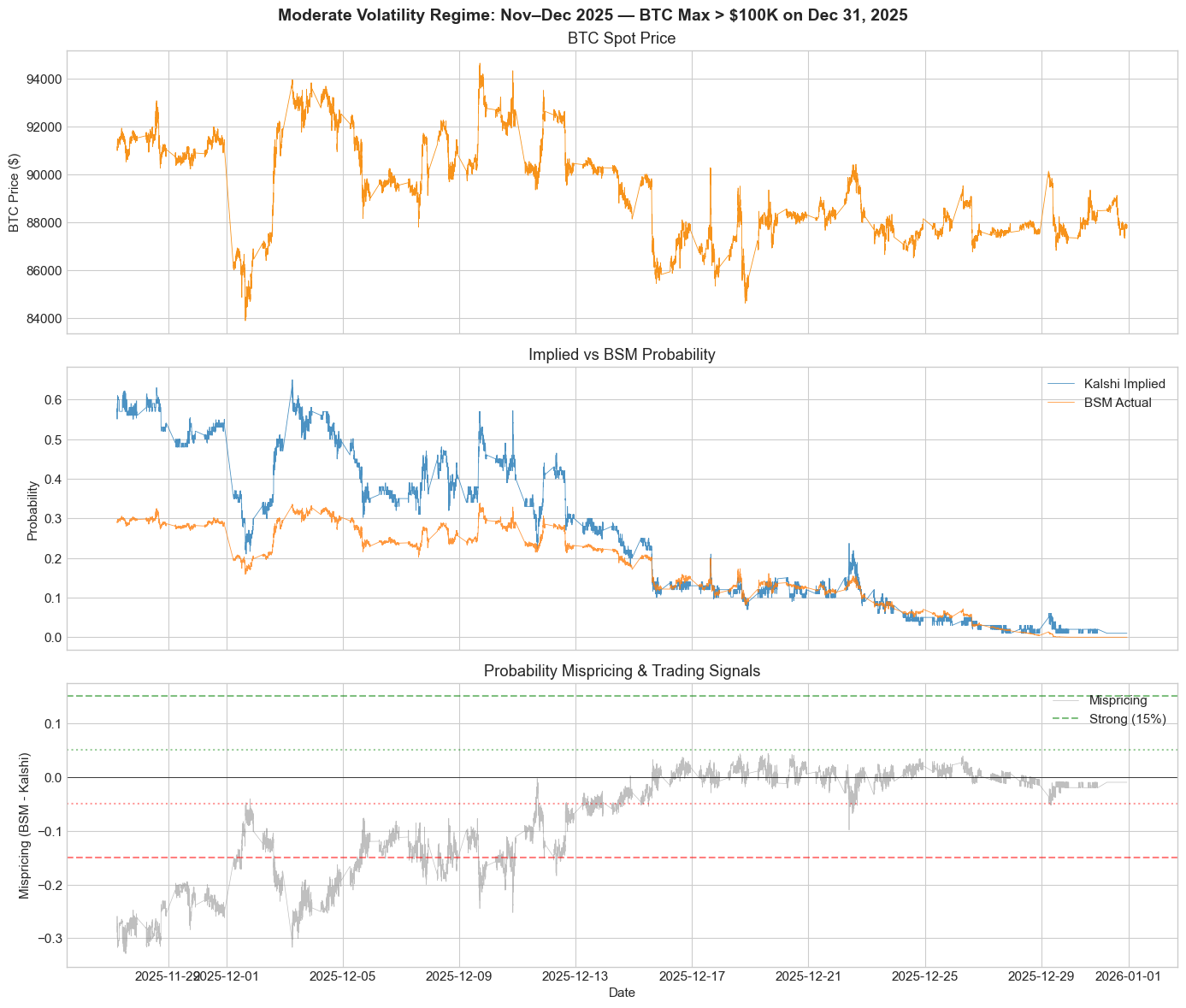}
  \caption{BTC Spot Price for Moderate Regime, with generated signals for Strategy C.}
  \label{fig:myplot}
\end{figure}

\noindent Fig 1 visualizes Strategy C’s signal generation for the November-December 2025 Kalshi event contract, with buying/selling indicators and magnitude. Implied and BSM probabilities compress as the event reaches its expiry and public consensus begins to fall in line with theoretical valuations. Trading signals are computed every minute, but the portfolio only trades at the start of each rebalancing period. Between rebalancing dates the position is held constant from the last rebalance.

\subsection{Fees and transaction costs}
Kalshi fees are incorporated into performance calculations using the platform's published quadratic fee schedule:

\begin{equation}
\mathrm{Fee} = \lceil.07\, C\, P(1-P)\rceil,
\label{eq:fee}
\end{equation}

\noindent Where C = contract size, and P = entry price for a singular contract. Aggressive order quantities were utilized in our framework in order to minimize profit erosion from transaction costs, which are integrated into each of the performance calculations examined under \textbf{Results}.

\section{Results}
We deployed each strategy on our 3 Kalshi markets, examining Total P\&L, percentage return, maximum drawdown and Sharpe ratio, focusing primarily on the dynamic hedging strategy’s performance relative to the unhedged baseline. The dynamic hedging portfolio was rebalanced on various different time periods, with 1 hour rebalances generally performing better than other time horizons. Our provided analysis therefore exclusively utilizes 1 hour rebalancing portfolios for comparison. Additionally, though we ultimately report results with fee impact taken into consideration, pure opportunity alpha is provided as well, since Kalshi fees are subject to regulatory change in the future.

\begin{figure}[htbp]
  \centering
  \includegraphics[width=0.9\textwidth]{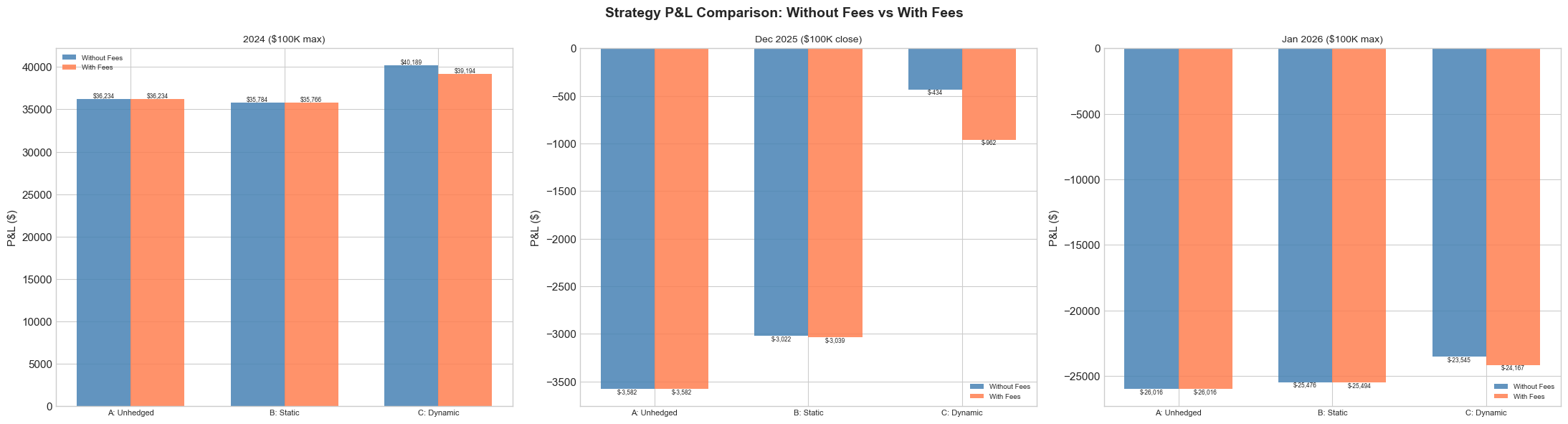}
  \caption{P\&L Charts with Fees Included for Strategies A,B,C across Kalshi Event Contracts.}
    \includegraphics[width=0.9\textwidth]{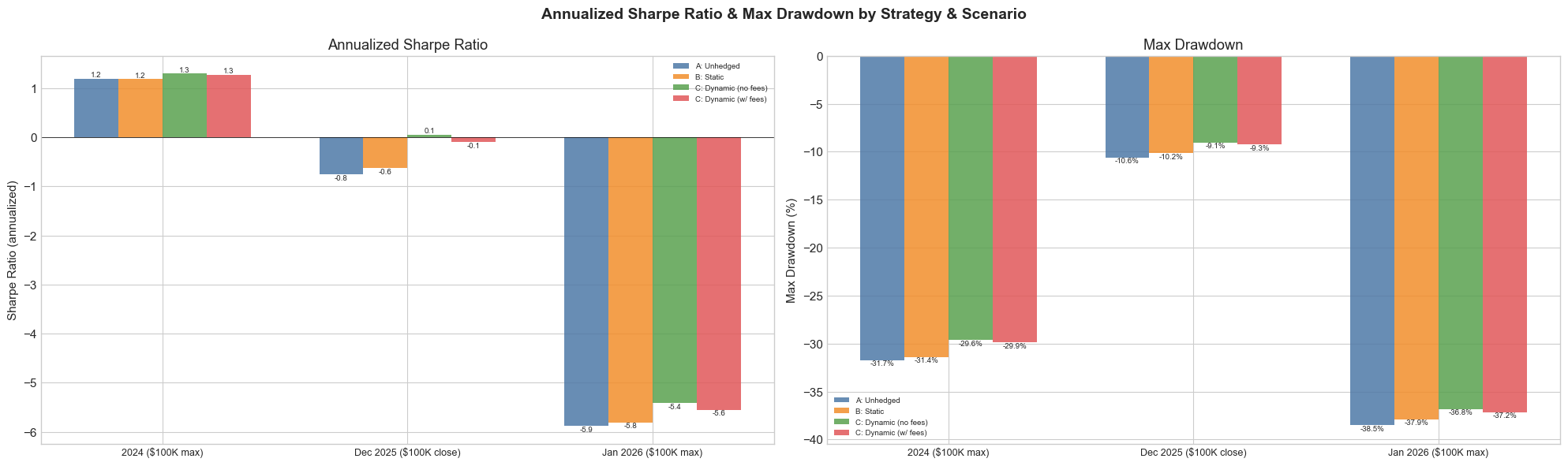}
  \caption{Sharpe/Drawdown Charts with Fees Included for Strategies A,B,C across Kalshi Event Contracts.}
  \label{fig:myplot}
\end{figure}
\newpage
\noindent Figures 2 and 3 showcase portfolio performance, focusing on P\&L, Sharpe Ratio, and Maximum Drawdown. Strategy C consistently outperforms strategies A and B, indicating hedging effectiveness during different market regimes. Outperformance in the bullish regime can be credited to  1 hour rebalancing periods, which allow the dynamic hedging portfolio to capture sudden market volatility and sentiment deviation.
\\\\
{\setlength{\textfloatsep}{8pt}\setlength{\intextsep}{8pt}\setlength{\floatsep}{8pt}

\begin{table}[htbp]
\centering
\small
\setlength{\tabcolsep}{6pt}
\renewcommand{\arraystretch}{1.1}
\begin{tabular}{lrr}
\hline
Market regime & BTC spot (Strategy A) & Dynamic hedge (Strategy C) \\
\hline
Bullish (2024) & +36.23\% & +38.69\% \\
Moderate/flat (Dec 2025) & $-3.58\%$ & $-3.04\%$ \\
Bearish (Jan 2026) & $-26.02\%$ & $-22.96\%$ \\
\hline
\end{tabular}
\vspace{0.005 in}
\caption{Spot vs.\ dynamically hedged returns across three distinct market regimes.}
\label{tab:regime_returns}
\end{table}

\begin{table}[htbp]
\centering
\small
\setlength{\tabcolsep}{4pt}
\renewcommand{\arraystretch}{1.08}

\resizebox{\textwidth}{!}{%
\begin{tabular}{p{0.22\textwidth} p{0.22\textwidth} r r r r r}
\hline
Scenario & Strategy & Sharpe & Max DD & VaR 95\% & CVaR 95\% & Fee \% Gross\\
\hline
\multirow{4}{*}{2024 \$100K max/}
& A: Unhedged & 1.20 & $-31.73\%$ & $-0.9614\%$ & $-1.6060\%$ & ---\\
& B: Static & 1.19 & $-31.40\%$ & $-0.9576\%$ & $-1.6026\%$ & ---\\
& C: Dynamic (no fees) & 1.30 & $-29.65\%$ & $-0.9415\%$ & $-1.5690\%$ & ---\\
& C: Dynamic (w/ fees) & 1.27 & $-29.90\%$ & $-0.9513\%$ & $-1.5802\%$ & 25.2\%\\
\hline
\multirow{4}{*}{Dec '25 \$100K max}
& A: Unhedged & $-0.76$ & $-10.59\%$ & $-0.7214\%$ & $-1.5052\%$ & ---\\
& B: Static & $-0.63$ & $-10.15\%$ & $-0.7116\%$ & $-1.4715\%$ & ---\\
& C: Dynamic (no fees) & 0.05 & $-9.06\%$ & $-0.7221\%$ & $-1.3953\%$ & ---\\
& C: Dynamic (w/ fees) & $-0.10$ & $-9.26\%$ & $-0.7249\%$ & $-1.4004\%$ & 16.8\%\\
\hline
\multirow{4}{*}{Jan '26 \$100K max}
& A: Unhedged & $-5.88$ & $-38.51\%$ & $-0.9892\%$ & $-1.6266\%$ & ---\\
& B: Static & $-5.81$ & $-37.89\%$ & $-0.9669\%$ & $-1.6093\%$ & ---\\
& C: Dynamic (no fees) & $-5.42$ & $-36.81\%$ & $-0.9499\%$ & $-1.5663\%$ & ---\\
& C: Dynamic (w/ fees) & $-5.55$ & $-37.17\%$ & $-0.9554\%$ & $-1.5778\%$ & 25.2\%\\
\hline
\end{tabular}%
}
\vspace{0.005 in}
\caption{\parbox{\textwidth}{Performance and risk metrics by scenario (Kalshi Market) and strategy.}}
\label{tab:scenario_strategy_metrics}
\end{table}
}
\noindent Tables 1 and 2 provide more detailed performance evaluations on each strategy’s performance per regime. We once again note that Strategy C outperforms A and B across all noted risk metrics, even when taking Kalshi fees into account. In each of the three examined regimes, Strategy C achieved roughly a 3\% reduction in exposure relative to the unhedged BTC position. In 2024 (BTC +36.23\%), the dynamic hedge returned 38.69\%; in December 2025 (BTC -3.58\%), strategy C returned -0.96\%; in January 2026 (BTC - 26.02\%) strategy C returned -24.17\%. Moreover, Strategy C exhibits pronounced improvements in Sharpe ratio over the unhedged portfolio, and lower expected Value at Risk at high thresholds.
\\
\\ However, the given results stress the importance of fee consideration. Across all regimes, roughly 20\% of Strategy C’s gains are eroded by Kalshi trading transactions, indicating that such fees make high-frequency mispricing arbitrage non-viable. Initial deployment of a strategy with hundreds of trades saw complete erosion of PL, implying that increased trading activity requires cost-benefit analysis to ensure that the expected profit of a given trade outweighs the cost of 7\% of its notional value. Additionally, as was mentioned before, further extensions must take into account latency and liquidity concerns in order to avoid unexpected slippage when executing orders to hedge exposure.

\section{Conclusion}
This paper showcased how Kalshi prediction market contracts can be utilized to reduce exposure for BTC investment strategies due to inherent differences between market-implied odds (sentiment-driven) and theoretical pricing. Our dynamic hedging strategy which actively rebalances binary outcome contract portfolios outperformed unhedged long and static-hedged long strategies in bull, bear, and flat markets. We conclude that Kalshi (and prediction markets in general) are useful risk management tools for crypto investors. However, scalability is still limited by liquidity, and future research should examine optimal rebalancing frequencies, alternative probability models, multi-strike portfolios, and different cryptocurrency prediction markets in order to maintain full control over a portfolio.

\section*{References}

\noindent [1] IG. "How to Hedge Bitcoin Risk" Accessed February 8, 2026. \\URL: \textbf{https://www.ig.com/en/trading-strategies/how-to-hedge-bitcoin-risk-190726}
\\\\
\noindent [2] Kalshi. “How high will Bitcoin get this year?” Accessed February 8, 2026. \\URL: \textbf{
https://kalshi.com/markets/kxbtcmaxy/how-high-will-bitcoin-get-this-year/btcmaxy-24dec31}
\\\\
\noindent [3] Kalshi. “Will Bitcoin cross \$100k again this year?” Accessed February 6, 2026. \\URL: \textbf{https://kalshi.com/markets/kxbtc2025100/will-bitcoin-reach-100k-again-this-year-/kxbtc2025100-25dec31}
\\\\
\noindent [4] Kalshi. “How high will Bitcoin get in January?” Accessed February 9, 2026. \\URL: \textbf{
https://kalshi.com/markets/kxbtcmaxmon/bitcoin-monthly-one-touch/kxbtcmaxmon-btc-26jan31}
\newpage
\section*{A. Additional Tables and Figures}
\addcontentsline{toc}{section}{Additional Tables and Figures}

\begin{figure}[htbp]
  \centering
  \includegraphics[width=1.0\textwidth]{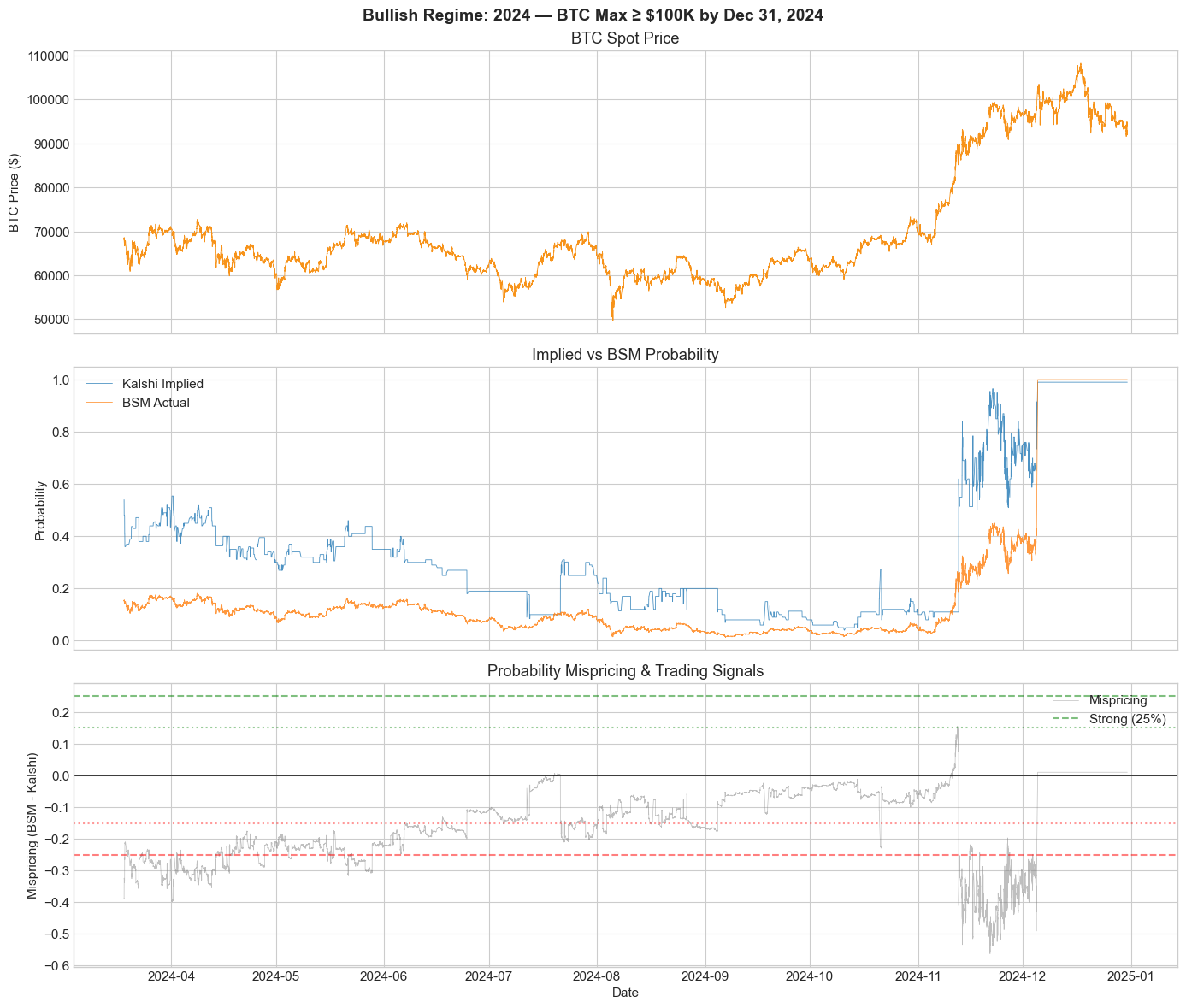}
  \caption{BTC Spot Price for Bullish Regime, with generated signals for Strategy C.}
  \label{fig:myplot}
\end{figure}

\begin{figure}[htbp]
  \centering
  \includegraphics[width=1.0\textwidth]{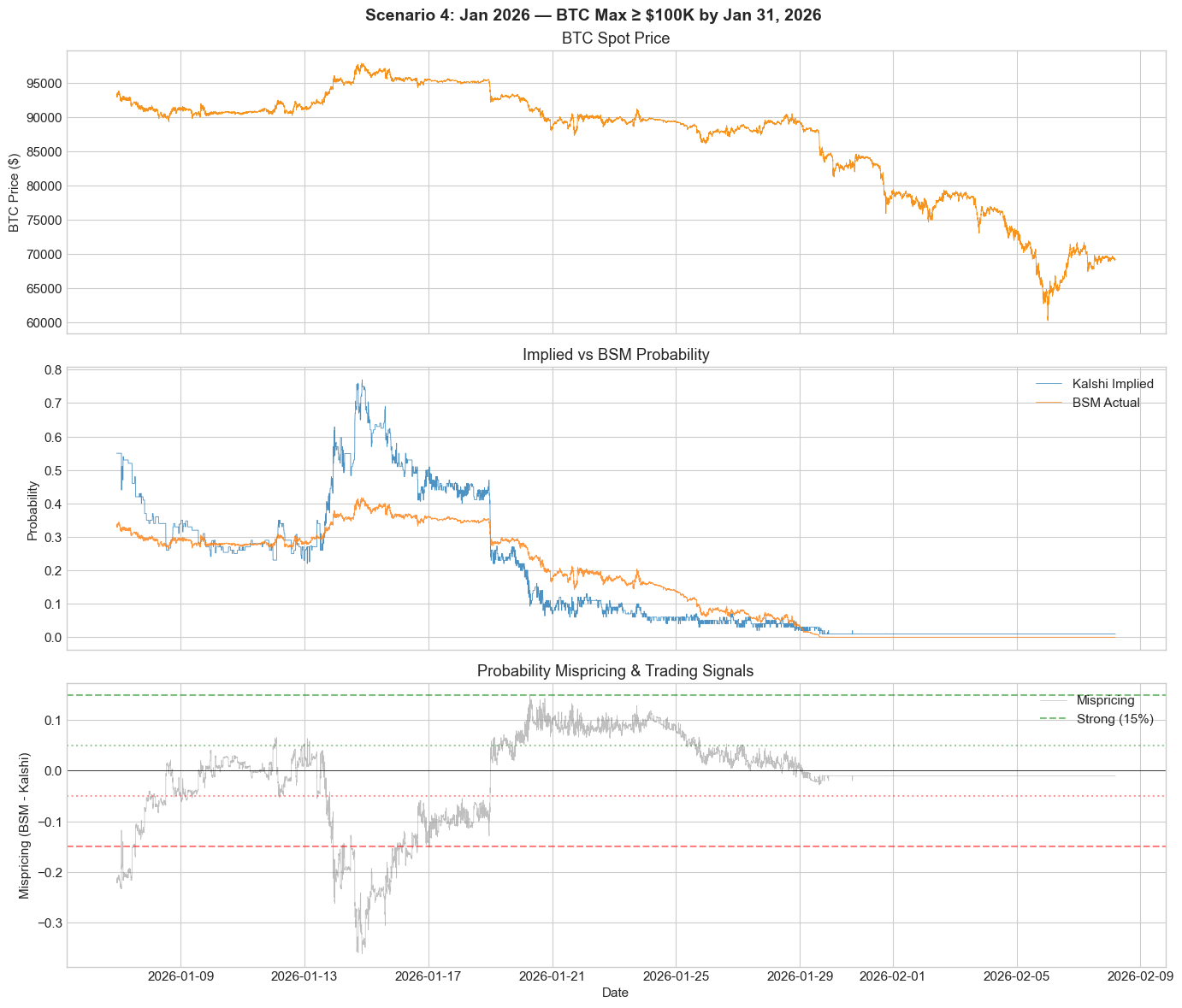}
  \caption{BTC Spot Price for Bearish Regime, with generated signals for Strategy C.}
  \label{fig:myplot}
\end{figure}

\begin{figure}[htbp]
  \centering
  \includegraphics[width=1.0\textwidth]{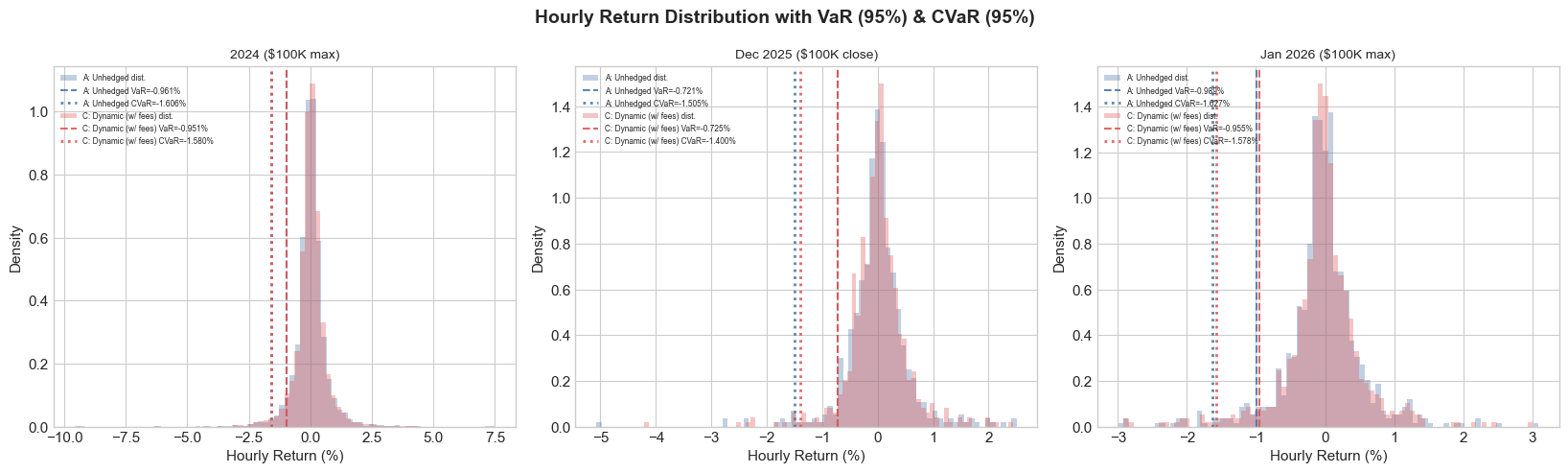}
  \caption{Value at Risk Visualizations utilizing Hourly Returns for all Strategies}
  \label{fig:myplot}
\end{figure}
\end{document}